\documentclass[numbook,epj]{svjour}
\usepackage{psfig}

\def\makeheadbox{
\rightline{cond-mat/9910318}
\rightline{ETH-TH/99-27}
}	

\def\Sv{\vec{S}}
\def\Sz{S^z}
\def\Sp{S^{+}}
\def\Sm{S^{-}}
\def\Jp{{J'}}
\def\abs#1{\vert #1 \vert}
\def\onehalf{{\textstyle {1 \over 2}}}
\def\Mag{\langle M \rangle}
\def\Order{{\cal O}}
\def\state#1{\vert #1 \rangle}
\def\astate#1{\langle #1 \vert}

\def\beq{\begin{equation}}
\def\eeq{\end{equation}}
\def\bea{\begin{eqnarray}}
\def\eea{\end{eqnarray}}
\def\nn{\nonumber}
\font\aefont=msam10
\def\age{\,\hbox{\aefont\char"26}\,}

\begin{document}

\title{A Spin-${1 \over 2}$ Model for CsCuCl$_3$ in an External Magnetic Field}

\author{A.\ Honecker\inst{1}\thanks{A Feodor-Lynen fellow of the Alexander
        von Humboldt-foundation}\thanks{Present address:
        Institut f\"ur Theoretische Physik,
        TU Braunschweig, Mendelssohnstr.\ 3, D--38106 Braunschweig,
        Germany}\and
        M.\ Kaulke\inst{2,3}\and
        K.D.\ Schotte\inst{2,4}}

\institute{Institut f\"ur Theoretische Physik, ETH-H\"onggerberg,
           CH--8093 Z\"urich, Switzerland,
           \email{a.honecker@tu-bs.de}
           \and
           Fachbereich Physik, Freie Universit\"at Berlin,
           Arnimallee 14, D--14195 Berlin, Germany
           \and
           \email{kaulke@physik.fu-berlin.de}
           \and
           \email{schotte@physik.fu-berlin.de}}

\date{October 19, 1999; revised December 28, 1999}

\abstract{CsCuCl$_3$ is a ferromagnetically stacked triangular
spin-1/2 antiferromagnet. We discuss models for its zero-temperature
magnetization process. The models range from three antiferromagnetically
coupled ferromagnetic chains to the full three-dimensional situation.
The situation with spin-1/2 is treated by expansions around the Ising
limit and exact
diagonalization. Further, weak-coupling perturbation theory is used
mainly for three coupled chains which are also investigated numerically
using the density-matrix renormalization group technique. We find that
already the three-chain model gives rise to the plateau-like feature
at one third of the saturation magnetization which is observed
in magnetization experiments on CsCuCl$_3$ for a magnetic field
perpendicular to the crystal axis.
For a magnetic field parallel to the crystal axis, a jump is observed
in the experimental magnetization curve in the region of again
about one third of the saturation magnetization.
In contrast to earlier spinwave computations, we do not find
any evidence for such a jump with the model in the appropriate
parameter region.}

\PACS{{75.10.Jm}{Quantized spin models} \and
      {75.45.+j}{Macroscopic quantum phenomena in magnetic systems} \and 
      {75.50.Ee}{Antiferromagnetics}
     }

\maketitle

\section{Introduction}

CsCuCl$_3$ is in several respects a quite unusual stacked triangular
(anti)ferromagnet (see e.g.\ \cite{CoPe} for a review of the subject).
Among others, three-dimensional features are vital in this compound:
The ferromagnetic stacking is just an order of magnitude larger than
the antiferromagnetic in-plane coupling while in most other materials
the coupling constants differ at least by two orders of magnitude.
Furthermore, the smaller antiferromagnetic coupling is crucial since it
gives rise to a non-trivial behaviour of CsCuCl$_3$ in the presence
of an external magnetic field.

The behaviour of CsCuCl$_3$ in strong external magnetic fields and
at low temperatures has been studied already some time ago
(see e.g.\ \cite{HTM}). One observes different behaviour,
depending on whether the magnetic field is applied along
or perpendicular to the $c$ (crystal)-axis. For a magnetic field
along the $c$-axis, there is a jump in the magnetization
curve at values of the magnetization of about one third
of the saturation value.
In contrast, one finds a plateau in the same area if the magnetic field
is applied perpendicular to the $c$-axis\footnote{Actually, a plateau with
$\Mag = 1/3$ can also be observed in other triangular
antiferromagnets (see e.g.\ \cite{IAG} for recent rather clear examples).}.

The spin in CsCuCl$_3$ is carried by Cu$^{2+}$ ions and thus
is a spin $1/2$. So, a proper treatment of quantum fluctuations
is important. However, in order to be able to treat the three-dimensional
situation, most theoretical approaches have so
far been either phenomenological \cite{StSu,NiJa,JaNi}
or used first-order spinwave theory \cite{NiShP,NiSh,JNS,RaTa,NiJa}.

The goal of the present paper is to treat directly a spin-$1/2$
model with realistic XXZ-anisotropies. We will use series
expansions around the Ising limit, numerical diagonalization and
degenerate second-order weak-coupling perturbation theory.

When one solves a classical or Ising-version of this model, it becomes
clear that essential features are already captured by a three-chain
variant. The bulk of our paper will therefore concentrate on three
antiferromagnetically coupled ferromagnetic chains. The simplification
to a one-dimensional situation also opens the possibility of
applying methods which work particularly well in one dimension like
the density-matrix renormalization group (DMRG) method \cite{White,DMRGbook}.

The purpose of the present paper is to
gain some qualitative insight and demonstrate that new theoretical
approaches to CsCuCl$_3$ are viable. It should be possible to develop
all these approaches further to improve the accuracy of the results,
to obtain a more realistic modeling or to compute other quantities.

\section{The model}

\begin{figure}[ht]
\centerline{\psfig{figure=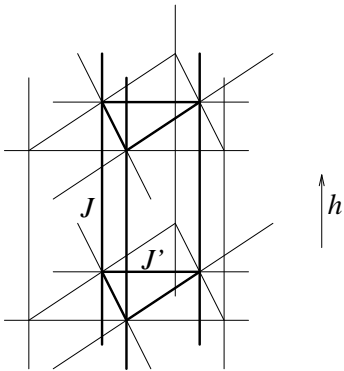,width=0.5\columnwidth}}
\smallskip
\caption{
The stacked triangular (anti)ferromagnet in a magnetic field $h$.
The bold lines denote our three-chain model.
\label{StackFig}
}
\end{figure}

Now we proceed to give a more accurate description of our model
and the parameter region we are interested in. We model CsCuCl$_3$
by a stacked triangular lattice antiferromagnet as sketched in
Fig.\ \ref{StackFig}.

The quantum-spin Hamiltonian corresponding to Fig.\ \ref{StackFig}
is given by (compare also \cite{NiSh}):
\bea
H &=& J \sum_{i=1}^N \sum_{x=1}^L
     \Delta_1 \Sz_{i,x} \Sz_{i,x+1} + \onehalf \left(
     \Sp_{i,x} \Sm_{i,x+1} + \Sm_{i,x} \Sp_{i,x+1} \right) \nn \\
  &&+\Jp \sum_{\langle i,j \rangle} \sum_{x=1}^L
     \Delta_2 \Sz_{i,x} \Sz_{j,x} + \onehalf \left(
     \Sp_{i,x} \Sm_{j,x} + \Sm_{i,x} \Sp_{j,x} \right)
\nn \\
  && - h \sum_{i,x} \Sz_{i,x} \, , \label{hamOp}
\eea
where the $\Sv_{i,x}$ are spin-$1/2$ operators and
$h$ is a dimensionless magnetic field. The notation
$\langle i,j \rangle$ denotes neighbouring pairs of the $N$ chains.
Here, the number of chains $N$ will be kept fixed while we are
interested in the limit of infinite $L$. Mostly, we choose
$\Delta_1 = \Delta_2$ and then denote both of them just by
`$\Delta$'.

Quite often, CsCuCl$_3$ is modeled by a Hamiltonian which includes
a Dzhalozhinski-Moriya interaction with a vector pointing along
the ferromagnetic chains in order to account for a modulation in
the magnetic structure with a period of slightly more than 70 layers
along the $c$-axis \cite{AAM}. However, by a local unitary transformation
this can be traded for a contribution to the anisotropy $\Delta_2$
and a surface term which is then discarded (see e.g.\ \cite{NiSh}).
Further contributions to the anisotropies come from crystal fields.
In this way, the $z$-direction in spin space is identified
with the direction along the ferromagnetic chains (the $c$-axis)
in real space.

$J$ will be ferromagnetic $(J < 0)$ and $\Jp$ antiferromagnetic $(\Jp > 0)$.
In order to describe the material CsCuCl$_3$ one uses $\Jp \approx -J/6$ and
$\Delta$ slightly less than one \cite{HIN,TTIN,TSS,MASOOYOY}. The actual values
used should not matter so much; it should however be noted that CsCuCl$_3$ is
described by weakly coupled ferromagnetic chains -- though not so weakly
coupled as is the case for many other triangular lattice antiferromagnets
\cite{CoPe}. A Monte-Carlo simulation of an XY--model takes $J' = -J/10$
\cite{PlMa} showing the influence the weaker coupling has on the critical
behaviour of triangular antiferromagnets. To make the situation clearer in
the study of the few-chain problem we choose a bit stronger interchain
coupling, e.g.\ $\Jp = -J/3$ to compensate partially for the missing
neighbouring chains.

The $\langle i,j \rangle$-summation in (\ref{hamOp}) should run
over a triangular lattice as sketched in Fig.\ \ref{StackFig}.
However, since $\Jp \ll \abs{J}$, a reasonable first approximation
should be obtained by retaining only one site as a representative
for each of the three sublattices of the triangular lattice.
This leads to the model of three coupled chains which is indicated
by the bold lines in Fig.\ \ref{StackFig}.

Since we are interested in the behaviour of CsCuCl$_3$ at low
temperatures, we actually simplify to zero temperature in
the present paper. We further restrict to magnetic fields applied along
the anisotropy axis. The magnetization 
\beq
\Mag = {2 \over N L} \left\langle
   \sum_{i=1}^N \sum_{x=1}^L \Sz_{i,x} \right\rangle
\label{defM}
\eeq
is then given as the expectation value of a conserved operator.
This is technically useful since it permits one to relate all
quantities in a magnetic field $h$ to those with a given magnetization
$\Mag$ at zero field. In the presence of an anisotropy $\Delta \ne 1$,
the effect of a magnetic field along the $z$-axis is different from
a magnetic field in the $xy$-plane. Because of the aforementioned
simplification we consider only the former case in the present
paper. 

Possible plateaux in the magnetization curve can be most easily read
off in the strong-coupling
limit $\Jp \gg \abs{J}$. For a fixed number $N$ of coupled chains
one finds \cite{CHP,CHPii} that for $J = 0$ the only possible
values of the magnetization are $\Mag = -1$, $-1+2/N$, ..., $1-2/N$, $1$.
In particular, for all $N$ that are divisible by three, a plateau with
$\Mag = 1/3$ is possible (the simplest case is $N=3$).
However, this strong-coupling argument does not ensure that
a certain plateau actually does occur for given values of the
parameters -- this issue requires a computation of the magnetization
curve (at least in the vicinity of the plateau-value of the
magnetization).

Before we proceed with the discussion of the magnetization process,
we note that a straightforward computation yields the upper critical
field for $N=3$ cyclically coupled chains
\beq
h_{uc} = \Jp (\Delta_2 + \onehalf) + J (\Delta_1 - 1)
\label{valHucChains}
\eeq
for $\Jp > 0$ and $J < 0$. This result is based on the assumption
that the transition proceeds with a single spin flip and is therefore
valid for $\Delta_1$ small enough (in particular $\Delta_1 \le 1$).
If the transition to saturation becomes first order,
(\ref{valHucChains}) ceases to be valid.

\section{Ising expansions for three chains}

The Ising limit of $N=3$ ferromagnetic chains (\ref{hamOp})
which are mutually antiferromagnetically coupled
is easily understood: At zero temperature, only
states with $\Mag = \pm 1$ and $\Mag = \pm 1/3$ are realized.
The groundstate at $\Mag = 1/3$
is given by a configuration in which all spins in two chains
point up and those in the third chain point down.
Perturbation expansions in $\Delta^{-1}$ around this state
and the ones with a single spin flipped respectively to
it are readily performed. One finds the lowest gaps for
the single spin excitations with states that are translational
invariant along the chain direction, and antisymmetrized between
the two equivalent chains for an additional spin flipped down.

With this information one proceeds in the same way as for
the triangular lattice \cite{twodim} (the precise method
has been summarized in \cite{Pert}) to find sixth-order
series for the gap of a single flipped spin. The series for
general $J$ and $\Jp$ can be found in appendix A. Here we
just present the specialization to $\Jp = -J/3$:
\bea
{h_{c_1} \over \abs{J}} &=& -\Delta +
{7 \over 6}-{\frac {11}{432 \Delta^{2}}}+{\frac {1993}{72576
\Delta^{3}}}-{\frac {5843545}{109734912 \Delta^{4}}} \nn \\
&& +{\frac {202825209149}{2824576634880 \Delta^{5}}}
   + {\cal O}(\Delta^{-6})
\, , \label{hc1ThreeChainSpec} \\
{h_{c_2} \over \abs{J}} &=&
{\frac {4 \Delta}{3}}-1-{\frac {1}{36 \Delta}}
+{\frac {11}{432 \Delta^{2}}}-{\frac {3887}{54432 \Delta^{3}}}
 \label{hc2ThreeChainSpec} \\
&& +{\frac {1558903}{18289152 \Delta^{4}}}-{\frac {
33098775077}{224682232320 \Delta^{5}}} + {\cal O}(\Delta^{-6})
\, . \nn
\eea
It should be noted that for sufficiently large $\Delta$ it is
actually favourable to flip large clusters of spins rather than
a single one if one changes the applied magnetic field.
Therefore, the transition at large $\Delta$ is
first order and (\ref{hc1ThreeChainSpec}) and (\ref{hc2ThreeChainSpec})
(or (\ref{hc2ThreeChain}) and (\ref{hc1ThreeChain}))
are not the boundaries of the $\Mag = 1/3$ plateau. Nevertheless, in
contrast to other Ising series such as those for the
triangular lattice \cite{twodim}, these series are remarkably well behaved
down into the region $\Delta \approx 1$ for the values of parameters we are
interested in ($\Jp$ small). The ending point of the $\Mag = 1/3$ plateau is
determined by the condition $h_{c_2}(\Delta_c) - h_{c_1}(\Delta_c) = 0$.
For example, for $\Jp/\abs{J} = 1/3$ we can use the raw series
(\ref{hc1ThreeChainSpec}) and (\ref{hc2ThreeChainSpec})
and find that the plateau ends at $\Delta_c =
0.965$, $0.876$ or $0.997$ if one truncates the series at
fourth, fifth or sixth order, respectively. At $\Jp/\abs{J} = 1/6$
convergence is a bit better and using (\ref{hc2ThreeChain}) and
(\ref{hc1ThreeChain}) one finds the ending point at
$\Delta_c = 0.972$, $0.954$ and $0.985$ for fourth, fifth
and sixth order, respectively.

This already indicates that the precise value of the XXZ-anisotropy
is crucial for explaining the presence or absence of a plateau with
$\Mag = 1/3$.

\section{Exact diagonalization for three chains}

\begin{figure}[ht]
\centerline{\psfig{figure=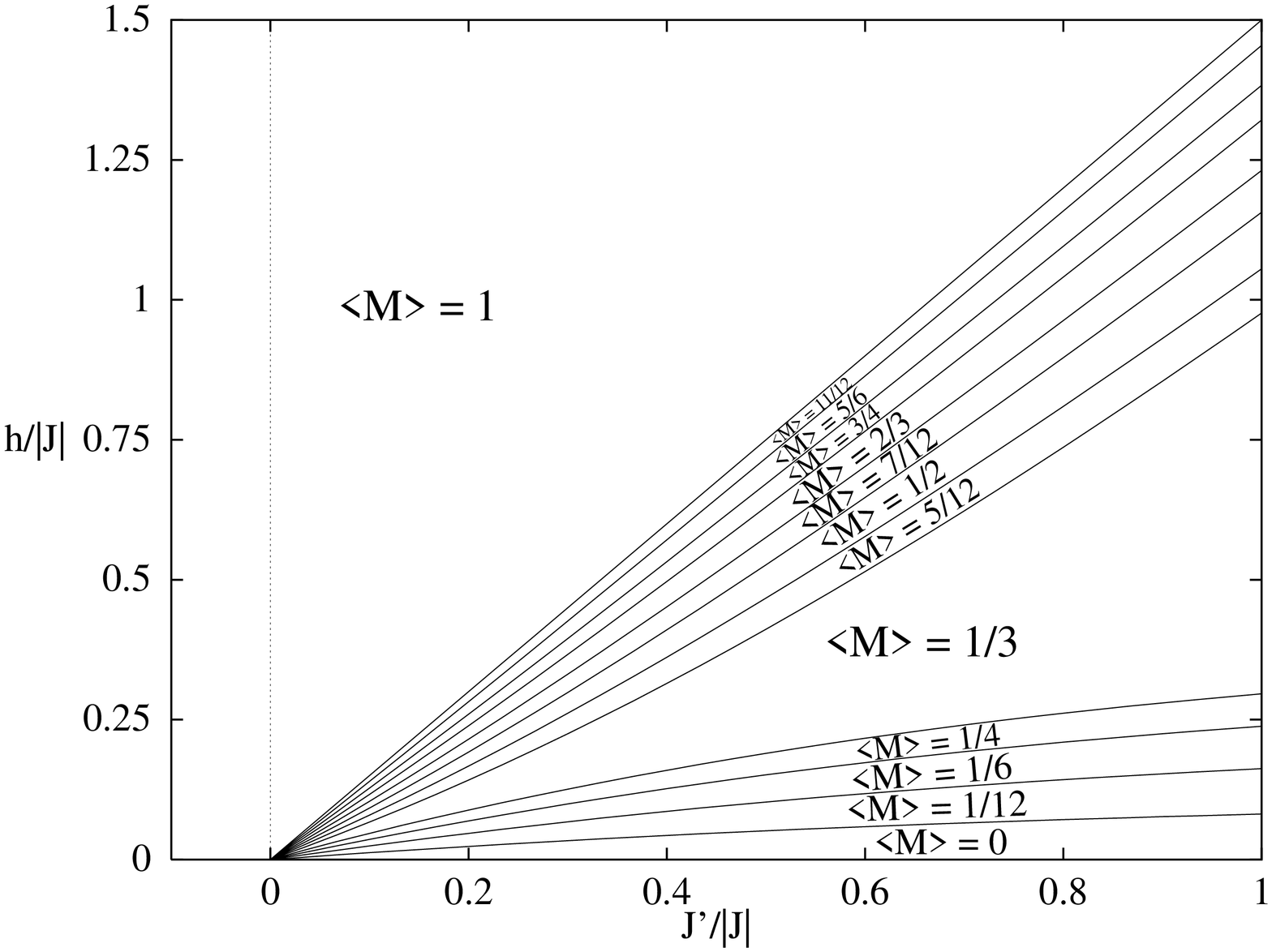,width=\columnwidth,angle=270}}
\smallskip
\caption{
Magnetic phase diagram for three coupled chains with $L=8$ and
$\Delta = 1$.
\label{FigJp}
}
\end{figure}

Next we present exact diagonalization results for three cylindrically
coupled chains with length $L=8$. Fig.\ \ref{FigJp} shows the magnetic phase
diagram for the $su(2)$-symmetric situation $\Delta = 1$.
The lines show the magnetic fields at which a transition
occurs between the states with the magnetization indicated in the
figure (for $L=8$, only $\Mag = m/12$ with $m=-12,\ldots,12$
can be realized).
One observes a plateau with $\Mag = 1/3$ for all $\Jp > 0$.
The transition to saturation clearly is given by
$h_{uc} = {3 \over 2} \Jp$,
as it should be according to (\ref{valHucChains}).
In the ferromagnetic regime $\Jp < 0$, the magnetization jumps
from $\Mag = -1$ to $\Mag = 1$ as $h$ passes through zero.

We would like to remark the striking smoothness of the transition
lines in the weak-coupling regime $\Jp \ll \abs{J}$ in Fig.\ \ref{FigJp}. This
suggests that expansions in $\Jp$ should be a viable way to understand
the magnetization process of three coupled chains (at least in a
situation with some $su(2)$ symmetry preserved). This will be the
subject of a later section.

For $\Jp \gg \abs{J}$, we can use the effective
Hamiltonian given in \cite{CHP} to determine the lower boundary
of the $\Mag = 1/3$ plateau. By exact
diagonalization we find with $\Delta = 1$ that $h_{c_1}/\abs{J} =
0.42096$, $0.41880$, $0.41779$, $0.41723$, $0.41689$
for $L=8,10,12,14,16$, respectively, and in the limit $J/\Jp \to 0$.
By comparison with Fig.\ \ref{FigJp} we see that the lower boundary
of the $\Mag = 1/3$ plateau must therefore still
increase beyond the right boundary of the figure.

\begin{figure}[ht]
\centerline{\psfig{figure=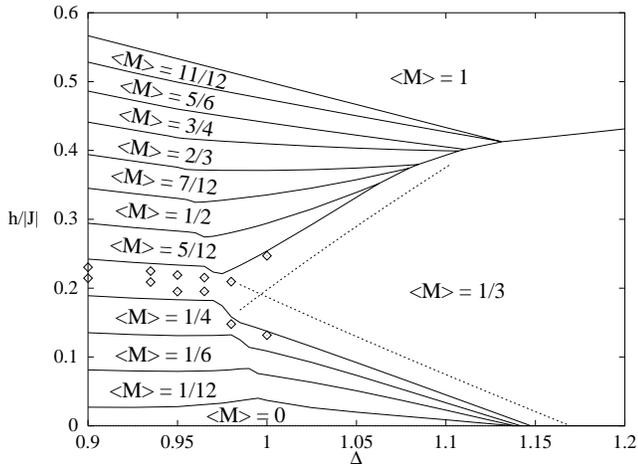,width=\columnwidth,angle=270}}
\smallskip
\caption{
Magnetic phase diagram for three coupled chains with $L=8$ and
$\Jp = \abs{J}/3$.
The dashed lines denote the sixth-order series (\ref{hc1ThreeChainSpec})
and (\ref{hc2ThreeChainSpec}).
The diamonds denote extrapolated values of the boundary of the
$\Mag = 1/3$ plateau (see next section).
\label{FigDelta}
}
\end{figure}

Now we fix the coupling constants to $\Jp = \abs{J}/3$
and look at the dependence on
the XXZ-anisotropy $\Delta$. The resulting magnetic phase
diagram for $L=8$ is shown in Fig.\ \ref{FigDelta}.

One observes that for $\Delta \age 1.14$ only the values
$\Mag = \pm 1/3$, $\pm 1$ occur. The corresponding
plateaux are separated by first-order transitions in this region.
Outside this region, the lower boundary of the $\Mag = 1/3$
plateau and the transition to saturation seem to remain second-order
transitions. The transition to saturation is given by (\ref{valHucChains})
in the region where it is of second order.

The $\Mag = 1/3$ plateau can be observed in Fig.\ \ref{FigDelta} for
$\Delta \age 0.98$, {\it i.e.}\ it survives a small amount of XY-like
anisotropy. The transition at its upper boundary becomes first order
for weak Ising-like anisotropies. For $L=8$ flipping two spins is
favoured over flipping just one for $\Delta \age 1.05$. A guess about what
happens in the thermodynamic limit $L \to \infty$ is not obvious.
Furthermore, in the vicinity of $\Delta \approx 1.13$ there is some
evidence for a jump in the magnetization curve at $h=0$.

The dashed lines in Fig.\ \ref{FigDelta} show the series expansions
(\ref{hc1ThreeChainSpec}) and (\ref{hc2ThreeChainSpec}) in $\Delta^{-1}$.
They denote the energy-cost to flip a {\it single} spin above the
$\Mag = 1/3$ plateau and should therefore be compared to
the corresponding full lines. The agreement of these sixth-order series
with the numerical data is quite good in view of the fact that the
window shown in the figure is rather far from the regime $\Delta \to \infty$
where the series were derived. These series point to a closing
of the plateau in about the same region indicated by the numerical
data.

We conclude this section by noting that the transition to full magnetization
is compatible with the Dzhaparidze-Nersesyan--Pokrovsky-Talapov (DN-PT)
universality class \cite{DzNe,PoTa}. From a technical point of view, it
is important that conservation
of the magnetization $M$ reduces the dimension of the space of states
considerably such that we can apply the exact diagonalization technique
to large lattices. Actually, for many purposes it is sufficient
to consider only two-spin deviations \cite{HoPa} which can be
easily treated numerically for volumes of around $1000$ spins.
Nevertheless, for $J<0$ one observes much stronger crossover
effects than for $J > 0$ \cite{CHPii}.

\section{DMRG}

Because of the one-dimensional topology of the three-leg ladder we
have used DMRG \cite{White} (for a detailed review of the DMRG method see also
\cite{DMRGbook}) in order to determine more accurately the value
of the anisotropy $\Delta$ where the plateau in the magnetization
occurs. Similar DMRG computations for the magnetization process of
a three-chain model with $J > 0$ have been carried out in \cite{tandon,COAIQ}
(see also \cite{WFMK} for a DMRG study of other spin ladders in a magnetic
field). Details on our implementation of the DMRG procedure can be found in
appendix B.

To do the calculations, three spins ${\bf S}_{i,x}$, $i=1,2,3$ for
fixed $x$ (one triangle in Fig.\ \ref{StackFig}) were combined to one site. In
this way one ends up with a spin chain of $L$ sites. We have
calculated the $\Mag = 1/3$ plateau for ladders of 90 spins (30 sites
in the corresponding chain model) for different values of the
anisotropy $\Delta$.

\begin{figure}[ht]
\centerline{\psfig{figure=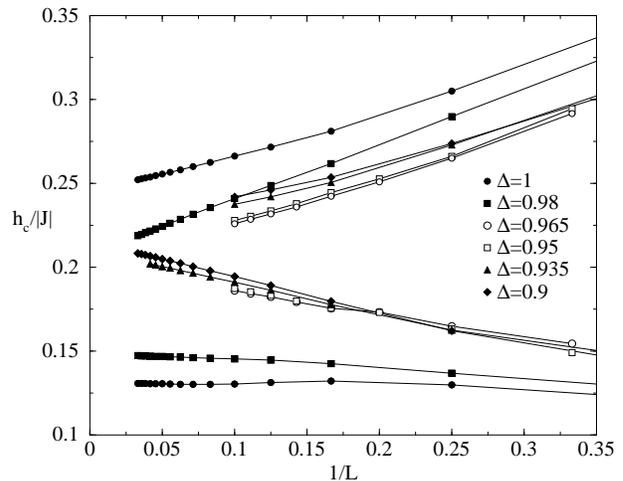,width=0.9\columnwidth}}
\smallskip
\caption{\label{fig_field} Upper and lower critical fields as a
function of $1/L$ for different values of $\Delta$. Filled symbols
denote DMRG results and open symbols data from exact
diagonalization.}
\end{figure}

Fig.\ \ref{fig_field} shows the upper and lower critical fields as
a function of the inverse chain length $1/L$. For large $\Delta$ $(\ge
0.98)$ the $\Mag = 1/3$ plateau
also exists in the thermodynamic limit. In the case of small $\Delta$
$(\le 0.935)$ the data is compatible with a vanishing plateau\footnote{
Variation of the truncation level in the DMRG procedure indicates
that the data may have errors up to $10^{-3}J\ldots10^{-2}J$ for the
larger system sizes.}. For $\Delta\le 0.8$ the plateau width
follows a $1/L$ law to a good approximation (this form
is a strong indication of a vanishing plateau width).

The interesting point is the threshold where the plateau
opens. However, the infinite-system algorithm of the DMRG method fails
already for short chain lengths when $\Delta$ approaches the region
around 0.95 where the threshold is expected. The reason for this is
not yet clear. Therefore in Fig.~\ref{fig_field} results from exact
diagonalization are plotted for these values of $\Delta$. The finite-size
data for the plateau-boundaries has been extrapolated to infinite $L$
using either a polynomial fit in $1/L$ or a fit with corrections of the
form $\exp(-c L)$. The former fit should work well for small $\Delta$
while the latter one is justified when a plateau is present, {\it i.e.}\
for the larger values of $\Delta$. The results of the extrapolation
are shown by the diamonds in Fig.\ \ref{FigDelta}. The error
of the extrapolation is smaller than the size of the symbols, but
there are additional systematic errors (e.g.\ in the DMRG truncation
procedure) which may be somewhat larger. Therefore, all the
data taken together indicates that the plateau closes indeed at
$\Delta \approx 0.95$ -- at least it becomes unobservable in this
region and for all practical purposes can be considered to be absent
for $\Delta < 0.95$.

\section{Weak-coupling expansions}

Since the coupling constants for CsCuCl$_3$ lie in the weak-coupling
regime $\Jp \ll \abs{J}$, it is natural to try to analyze (\ref{hamOp}) by
an expansion in $\Jp$ around the decoupled point $\Jp = 0$. Such
an approach is even more strongly suggested by Fig.\ \ref{FigJp} which shows
that at least for $N=3$ and $\Delta = 1$, the transition lines are
very smooth functions of $\Jp/\abs{J}$.

\subsection{First order}

We choose $\Delta_1 = 1$ in (\ref{hamOp}) for the weak-coupling expansions
because of two reasons ($\Delta_2$ will however be retained as a
parameter). Firstly, we can then exploit the $su(2)$ symmetry of the
decoupled chains in order to immediately write down their
groundstates for a given magnetization. Secondly, in this case
we can expect a smooth dependence on the coupling $\Jp$
(see Fig.\ \ref{FigJp}), which is not present otherwise (this is indicated
by numerical diagonalization -- compare also Fig.\ \ref{FigDelta}).
A consequence of this choice $\Delta_1 = 1$ is that a given
magnetization can be arbitrarily distributed among the individual
chains. This means that the groundstate is highly degenerate and
one has to perform degenerate perturbation theory.

The groundstate space of a single $su(2)$-symmetric ferromagnetic
Heisenberg chain of length $L$ is the spin-$L/2$ representation.
Thus, one can immediately write down the groundstate with
a given total ${\cal S}^z$ eigenvalue $m$:
\beq
\state{j=L/2,m} = {1 \over \sqrt{{\cal N}_{j,m}}}
\left({\cal S}^{-}\right)^{j-m} \state{j,j} \, ,
\label{SzEigen}
\eeq
where ${\cal S}^{-}$ is the total step operator for a single chain.
Here $j$ denotes this spin and $m$ the eigenvalue of the $z$-component
of the total spin operator. The normalization ${\cal N}_{j,m}$
in (\ref{SzEigen}) should be chosen such that all states $\state{j,m}$
are normalized to one. Otherwise it is sufficient to know that the
number of terms on the r.h.s.\ of (\ref{SzEigen}) is
${2 j \choose j-m}$.

Because of translational invariance one readily finds that
\bea
\astate{j,m} \Sz_{i,x} \state{j,m}
  = {1 \over L} \astate{j,m} {\cal S}^z \state{j,m}
  = {m \over L} = {m \over 2 j} \, .
\label{SzEigenVal}
\eea
A simple combinatorial consideration further leads to
\bea
\astate{j,m+1} \Sp_{i,x} \state{j,m}
&=& {\sqrt{(j-m) (j+m+1)} \over 2 j} \, , \nn \\
\astate{j,m-1} \Sm_{i,x} \state{j,m}
&=& {\sqrt{(j-m+1) (j+m)} \over 2 j} \, .
\label{SpmMatEl}
\eea
Note that due to translational
invariance the matrix elements (\ref{SzEigenVal}) and (\ref{SpmMatEl})
do not depend on the position of the spin-operator $x$. The
summation over $x$ in (\ref{hamOp}) therefore just yields a factor
$L = 2 j$.

In (\ref{SzEigenVal}) and (\ref{SpmMatEl}) one recognizes the
matrix elements of the $\vec{J}$ operators of a spin-$j$
$su(2)$-representation normalized by a factor $1/(2j)$. Thus, the
interaction part of the Hamiltonian (\ref{hamOp}) can be rewritten
in first order as
\beq
H_I = {\Jp \over 2 j} \sum_{\langle k,l \rangle}
   \Delta_2 J^z_k J^z_l + \onehalf \left(
   J^+_k J^-_l + J^-_k J^+_l \right)
   - h \sum_k J^z_k \, .
\label{hInt}
\eeq
This form is particularly useful for an analytical treatment. Firstly,
one sees that for $L=2j \to \infty$, the first-order interaction maps
to a problem of classical spins where each spin stands for one chain.
In the groundstate of a classical spin model on the triangular lattice,
all spins on one sublattice point in the same direction. Therefore,
a triangular arrangement of ferromagnetic chains and
three coupled chains become equivalent at first order in $\Jp$
-- only the coupling constant is rescaled by the different coordination
number.

Using this mapping to a classical spin model, a number of conclusions
can be drawn which apply to three coupled chains as well as the full
triangular arrangement:
\begin{enumerate}
\item The entire magnetization curve is linear at $\Delta_2=1$.
\item For $\Delta_2 > 1$, the magnetization jumps from
  $\Mag = -m_0$ to $\Mag = m_0$ with $m_0 = (\Delta_2 - 1)/
  (3 (\Delta_2 + 1))$ as $h$ passes through zero
  (see also Fig.\ \ref{FigDelta}).
\item For $\Delta_2 > 1$ there is a plateau with $\Mag=1/3$.
  Its lower boundary is given by $h_{c_1} = z \Jp/4$ where
  $z$ is the number of nearest neighbours ($z=2$ for three coupled
  chains and $z=6$ for an underlying triangular lattice). The upper boundary
  is more difficult to determine analytically -- a lower bound\footnote{
  The important point is that $h_{c_2} > h_{c_1}$ for
  $\Delta_2 > 1$.}
  is given by $h_{c_2} \ge z \Delta_2 \Jp/4$.
\item The asymptotic behaviour of $\Mag$ as a function of $h$
  is linear close to saturation or to the boundaries of the
  $\Mag = 1/3$ plateau:
\beq
\Mag - M_c \propto h - h_c \, .
\label{linAsym}
\eeq
\end{enumerate}
With $N=3$ chains, the first conclusion can actually be obtained for
finite $L$ using a different argument which is also useful for
determining the groundstate degeneracy:
For $\Delta_2=1$ and $N=3$ one can further rewrite eq.\ (\ref{hInt}) as
\bea
H_I &=& {\Jp \over 4 j} \left(\vec{J}_1 + \vec{J}_2 + \vec{J}_3\right)^2
 - {3 \Jp (j+1) \over 4} \nn \\
&& - h \left(J^z_1 + J^z_2 + J^z_3\right) \, .
\label{hIntIso}
\eea
This shows that the energy is a quadratic function of the total
spin (or the magnetization $\Mag$). Therefore, all steps in the finite-size
magnetization curve acquire equal width at $\Delta_2 = 1$.
Also the groundstate degeneracy can
be easily inferred from the representation (\ref{hIntIso}). It is
found to be maximal at $\Mag = 1/3$ where it is equal to $L+1$
and minimal at $\Mag=0,1$ where the groundstate is unique.

The treatment of the second order is going to be based on a numerical
determination of the groundstate of the interaction $H_I$ at $N=3$ and finite
$L$. It is then important that the groundstate should be non-degenerate
at first order. This precludes an analysis of the second order for
the case $\Delta_2 = 1$ where after treatment of the first order one
is still left with large degeneracies. For $\Delta_2 \ne 1$,
the groundstates with a given magnetization are at most twofold
degenerate\footnote{Still, for $\Delta_2 > 1$ there are excited states above
the groundstate(s) for a given magnetization which have a gap
which closes rapidly with increasing system size $L$. Therefore, depending
on the chain length $L$, the magnetization and the numerical
accuracy, groundstate degeneracies may appear to be threefold
in the region $\Delta_2 > 1$.}. This twofold degeneracy is due to
the spatial symmetries of the three-chain model which can be used
to lift it completely (alternatively, such a two-fold degeneracy
can be ignored since it does not affect the results). Thus, in particular
the case we are mainly interested in, namely $\Delta_2 < 1$ is amenable
to a numerical treatment up to the second order.

\subsection{Second order}

If one goes beyond the first order, one has to consider also
excitations that are created by the interaction
from the ferromagnetic groundstates (\ref{SzEigen}). At
second order, only a single spin can be flipped in each chain.
To deal with the second order perturbation it is therefore
sufficient to look only at spinwave states
\beq
\state{j,m,k} = {1 \over \sqrt{\hat{\cal N}_{j,m}}} \sum_{x = 1}^L
      {\rm e}^{i k x} \Sm_{i,x} \state{j,m+1}
\label{swState}
\eeq
with a suitable normalization factor $\hat{\cal N}_{j,m}$.
This representation is useful because the states (\ref{swState})
are eigenstates of the Hamiltonian $H_0$ for a single chain
\beq
\left(H_0 - {J L \over 4}\right) \state{j,m,k}
   = \abs{J} \left(1 - \cos\left(k\right)\right) \state{j,m,k} \, .
\label{swEn}
\eeq
The computation of the matrix elements is now a bit more
cumbersome than for the first order but follows the same lines
of combinatorial considerations. In this manner one
finds the following counterpart of (\ref{SzEigenVal}) for $k \ne 0$: 
\beq
\astate{j,m,k} \Sz_{i,x} \state{j,m,0}
  = - {{\rm e}^{- i k x} \over 2 j} \sqrt{{(j-m) (j+m) \over 2 j - 1}} \, .
\label{SzMelSw}
\eeq
The $k\ne 0$ generalization of (\ref{SpmMatEl}) is found to be
given by
\bea
\astate{j,m+1,k} \Sp_{i,x} \state{j,m}
&=& - {{\rm e}^{- i k x} \over 2 j}
  \sqrt{{(j-m-1) (j-m)} \over 2 j-1} \, , \nn \\
\astate{j,m-1,k} \Sm_{i,x} \state{j,m}
&=& {{\rm e}^{- i k x} \over 2 j}
  \sqrt{{(j+m-1) (j+m)} \over 2 j-1} \, . \nn \\
\label{SpmMelSw}
\eea
With the matrix elements (\ref{SzMelSw}) and (\ref{SpmMelSw})
one can now easily write down the matrix elements for the
interaction between two chains. It should be noted that due
to translational invariance, spinwave states with momenta
$k$ and $-k$ have to come in pairs (in other words:
$\sum_x {\rm e}^{- i (k+k') x} = L \delta_{k+k',0}$).
This leads to a cancellation of the phase factors in (\ref{SzMelSw}) and
(\ref{SpmMelSw}) and thus makes the matrix elements of the interaction
$k$-independent. Then the $k$-summation can be pulled out from
the second-order matrix. So, one obtains for the second-order
contribution
\beq
\astate{\psi_1} V \left({J L N \over 4} - H_0 \right)^{-1} V \state{\psi_1}
= {\cal E}^{-1} \astate{\psi_1} H_{II}^\dagger H_{II} \state{\psi_1} \, ,
\label{2ord}
\eeq
where $\state{\psi_1}$ is the eigenstate of the first-order matrix
$H_I$. The energy denominator is given by
\beq
{\cal E}^{-1} = - {1 \over 2 \abs{J}} \sum_{\kappa=1}^{L-1}
    {1 \over 1 - \cos\left({2 \pi \kappa \over L} \right)} 
    = - {\left(L-1\right) \left(L+1\right) \over 12 \abs{J}} \, .
\label{eDen}
\eeq
In order to write $H_{II}$ in a compact form,
we introduce operators $E^z_i$, $E^\pm_i$ acting in the groundstate space
of the $i$th chain as
\beq
E^z_i \state{j,m} = \state{j,m} \, , \qquad
E^\pm_i \state{j,m} = \state{j,m\pm1} \, .
\label{defEop}
\eeq
Since the matrix elements of $H_{II}$ do not depend on $k$
one can identify all spaces
spanned by $\state{j,m,k}$ for a given $k$ with that spanned by
$\state{j,m}$. Then $H_{II}$ can be written as
\bea
{H_{II} \over \Jp} &=& \sum_{\langle i,l \rangle} \Biggl\{ \Delta_2
   {\sqrt{(j-m_i) (j+m_i) (j-m_l) (j+m_l)} \over 2 j ( 2 j - 1)}
   E^z_i E^z_l \nn \\
&& - {\sqrt{(j-m_i-1) (j-m_i) (j+m_l-1) (j+m_l)} \over 4 j ( 2 j - 1)}
   E^+_i E^-_l \nn \\
&& - {\sqrt{(j+m_i-1) (j+m_i) (j-m_l-1) (j-m_l)} \over 4 j ( 2 j - 1)}
   E^-_i E^+_l \Biggr\} \, . \nn \\
\label{h2ord}
\eea
Here the $m_i$ denote the quantum numbers of the state on which $H_{II}$
is acting.
The combination $H_{II}^{\dagger} H_{II}$ appearing in (\ref{2ord}) can
be expressed by spin operators. In the isotropic case $\Delta_2 = 1$
a simple form can be found
\beq
{H_{II}^{\dagger} H_{II} \over \Jp^2} = \sum_{\langle i,l \rangle}
{(\vec{J}_i \vec{J}_l)^2 - 2 j(j - 1) \vec{J}_i \vec{J}_l + j^3 (j-2)
\over 4j^2 (2j -1)^2}
\label{h2ordIso}
\eeq
which reduces for large chain length $L = 2j \gg 1$ to
$\sum_{\langle i,l \rangle} (\vec{J}_i \vec{J}_l/j^2 - 1)^2/16$
that is to the exchange
between two chains squared instead of the simple exchange in first order
perturbation theory given by (\ref{hInt}). The squared exchange has been used
as a tool to take into account quantum or thermal fluctuations (see
\cite{NiJa} and references therein).

So far, everything is valid for a general underlying lattice. Let
us now comment on the results obtained from this second-order
approach for three chains. For this case, one can easily determine
the groundstate $\state{\psi_1}$ of (\ref{hInt}) numerically for
$L$ up to a few hundred when $\Delta_2 < 1$. For short chains,
the first and second order expansion terms obtained from (\ref{hInt})
and (\ref{2ord}) compare favourably with a direct diagonalization
of the Hamiltonian (\ref{hamOp}). However, the second-order corrections
to the energy obtained from (\ref{2ord}) behave as $L^2$ for large $L$.
Alternatively, the second-order
corrections to $h(\Mag)$ computed from these energies are proportional
to $L$ for long chains. This means that the radius of convergence shrinks
rapidly to zero with increasing $L$. The additional factor $L$ which
limits the second-order weak-coupling approach for three coupled
chains can be traced to the energy denominator (\ref{eDen}) which is
the main source of the system-size dependence.

With hindsight this divergence of the second order can be expected
since a one-dimensional ferromagnet cannot be stable to a weak
antiferromagnetic perturbation if the chain length is long. Second order
perturbation theory connects only two chains, if one compares with
contributions to the ground state energy $\propto \Jp^{3/2}$ one gets from a
Holstein--Primakoff analysis for two ferromagnetic chains coupled
antiferromagnetically one notices immediately the difficulties.
Further in the limit of vanishing $\Jp$, one would find a linear behaviour of
the magnetization (\ref{linAsym}) in the thermodynmic limit. Now, if one takes
this limit $L \rightarrow \infty$ at fixed $\Jp > 0$, the asymptotic DN-PT
square root behaviour$\Mag - M_c \propto \sqrt{\abs{h - h_c}}$ which is
characteristic for one dimension \cite{DzNe,PoTa} has to be recovered. Such
an abrupt change in the functional form is simply not compatible whith a
convergent perturbation expansion in the thermodynmic limit.
Only in three and higher dimensions, the functional behaviour (\ref{linAsym})
remains valid for the interacting quantum system (see section 7 of \cite{twodim}
and references therein). So, the second-order correction has a chance
to be convergent only if the thermodynamic limit is taken in all
three directions simultaneously. On the one hand, this is the case
most relevant to CsCuCl$_3$. On the other hand, already determination
of the first order groundstate involves the solution of a spin-$L/2$
problem on the triangular lattice. Then the numerical effort becomes
much larger than in the three-chain model and one gains little by
the perturbative treatment in comparison to a direct numerical treatment
of the full problem (which will be carried out in section 8.2).
Therefore, here we do not pursue evaluation of the second order for
the full three-dimensional situation further.

\section{Exact diagonalization results for six chains}

Now we briefly look at the Hamiltonian (\ref{hamOp}) for
$N=6$ chains. The coupling of the nearest neighbour pairs
$\langle i,j \rangle$ ($1 \le i,j \le 6$) is specified by
the lines in the following picture:
\beq
\lower1.6cm\hbox{
\psfig{file=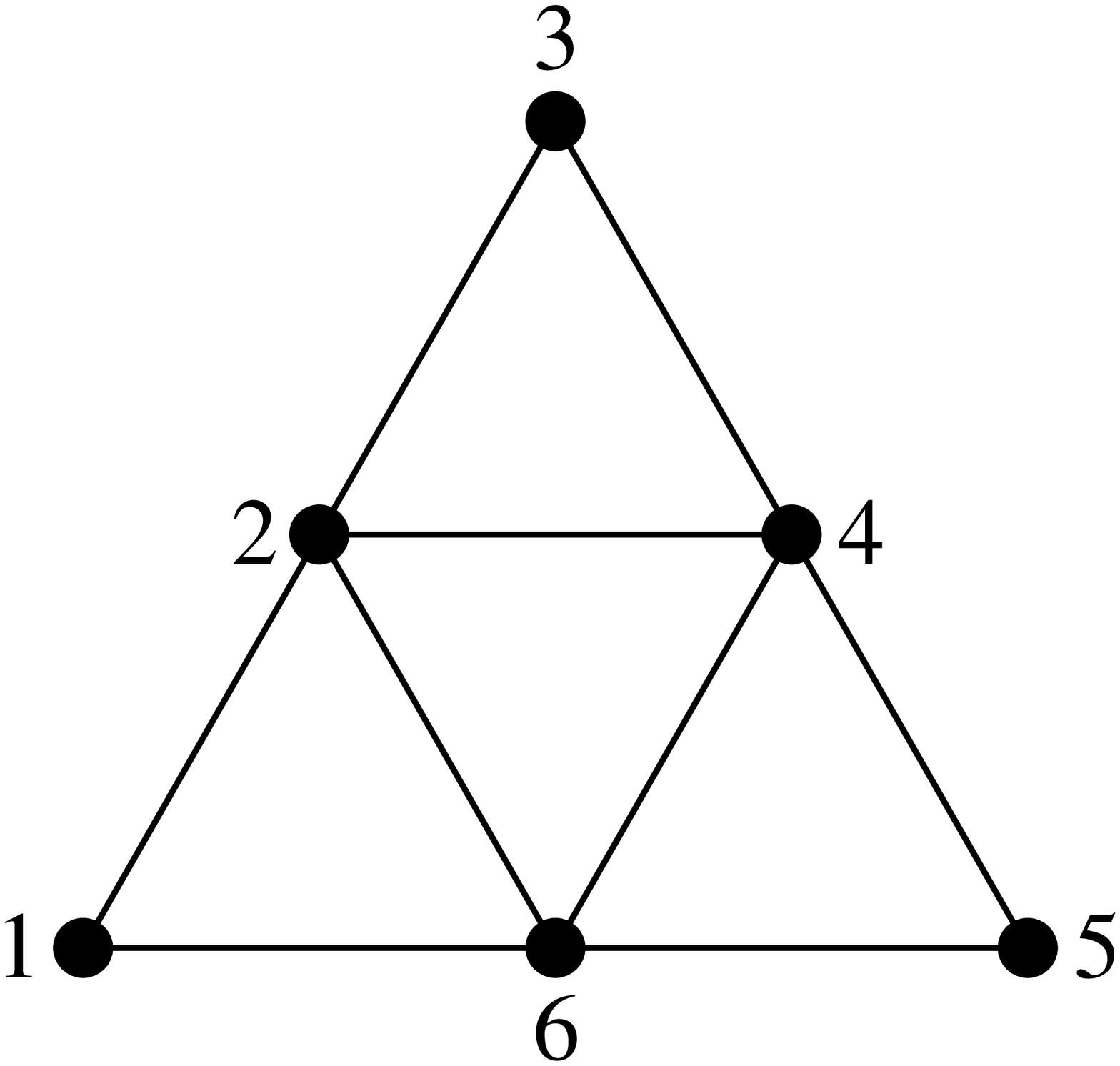,width=0.4\columnwidth}}
\label{SixCoupl}
\eeq
This is basically a refined version of the $N=3$ chain model
discussed in the preceding sections and thus a natural step
in the direction of the three-dimensional compound. In particular,
one can see that $N$ must be a multiple of three if one wishes
to expect a plateau with $\Mag = 1/3$ just on the
basis of counting the chains. For $N=6$, this counting argument
indicates the possibility of having plateaux with $\Mag \in \{0,
\pm 1/3, \pm 2/3, \pm 1 \}$. If one now diagonalizes a single
layer ($L=1$) of the model specified by (\ref{hamOp}) and
(\ref{SixCoupl}), one finds that all these values are indeed
realized one after the other in the presence of a magnetic
field.

\begin{figure}[ht]
\centerline{\psfig{figure=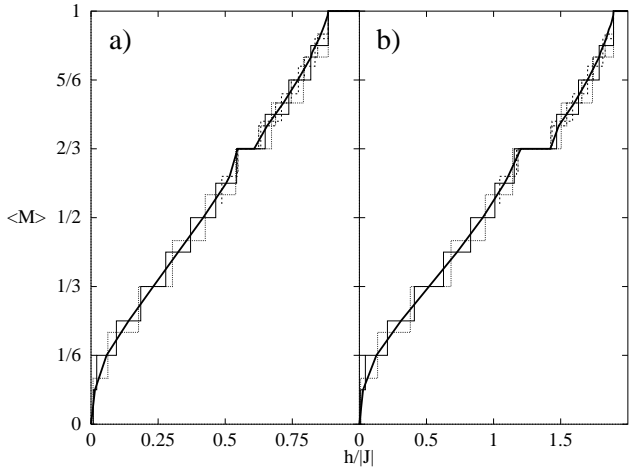,width=\columnwidth,angle=270}}
\smallskip
\caption{
Magnetization curves for six coupled chains with $\Delta = 1$
and a) $\Jp/\abs{J} = 1/3$, b) $\Jp/\abs{J} = 5/7$.
The number of layers is $L=3$ (dotted line), $L=4$
(thin full line), $L=5$ (dashed line) and $L=6$ (dashed-dotted line).
The thick full line is the extrapolation to $L = \infty$
discussed in the text.
\label{sixMag}
}
\end{figure}

In a way similar to section 4 we have computed the
complete magnetization curve of this six-chain model
for $L \le 4$ layers and parts of this curve for $L \le 35$
at two values $\Jp/\abs{J} = 1/3$, $5/7$ and $\Delta = 1$
\footnote{The average coordination in the
six-chain model (\ref{SixCoupl}) is larger than in the three-chain
model which suggests to also investigate smaller $\Jp$. However,
the data is already somewhat ambiguous at $\Jp/\abs{J} = 1/3$.
Therefore, we have added a larger value of $\Jp$ rather than a smaller one.}
The results for $3 \le L \le 6$ are shown in Fig.\ \ref{sixMag}. The
thick full lines in this figure are somewhat speculative
extrapolations to $L = \infty$ which need discussion.

For all the sizes which we have investigated, there is an $\Mag = 0$
plateau (or equivalently a spin-gap), which however
vanishes rapidly with increasing $L$. We are therefore confident
that there is no $\Mag = 0$ plateau in the thermodynamic limit
of the six-chain model, but rather a steep increase around
$\Mag = 0$ as indicated by the thick full line in Fig.~\ref{sixMag}.

To see what happens to the other two plateaux which are
permitted according to the aforementioned counting argument
({\it i.e.}\ $\Mag = 1/3$ and $\Mag = 2/3$)
we have performed extrapolations of the available
data for the boundaries of these plateaux using the vanden
Broeck--Schwartz algorithm (see e.g.\ \cite{HeSchue}).
This works better for $\Mag = 2/3$ where we
can use data for $2 \le L \le 6$ than for $\Mag = 1/3$
where just $L=2,3,4$ is accessible. The result for $\Mag = 1/3$
is compatible with a plateau of zero width while for $\Mag = 2/3$
the error estimates indicate a non-vanishing plateau. The conclusion
of the $\Mag = 1/3$ plateau being absent in the thermodynamic
limit is further supported by the observation that at finite size
this step is about as wide as the neighbouring ones. On the other hand,
the width of the $\Mag = 2/3$ step is notably larger than
that of its neighbours -- at least for $\Jp / \abs{J} = 5/7$.

Therefore we have decided to draw plateaux at $\Mag = 2/3$
in Fig.\ \ref{sixMag} but none for $\Mag = 1/3$. The boundaries of
the former case are the only points where fairly reliable extrapolations
to $L = \infty$ are available. Otherwise we follow the procedure
of \cite{BoFi,PaBo} and draw a line through the midpoints of the
steps at the largest available system size, although it is apparent
from Fig.\ \ref{sixMag} that finite-size effects are still of some
importance.

Finally, the transition to saturation is of the same type as that
for the three-leg ladder, {\it i.e.}\ it is again compatible with
the DN-PT universality class \cite{DzNe,PoTa}, but subject to
strong crossover effects in the region $J < 0$.

The main result of this discussion of $N=6$ chains is
that already doubling the number of chains leads to a vanishing of
the $\Mag = 1/3$ plateau -- at least for $\Delta = 1$ and weak
coupling $\Jp$. This is a first indication that this plateau which
is easily observable in $N=3$ coupled chains might actually not
survive the infinite-lattice limit in the plane perpendicular to
the chains.

\section{The three-dimensional case}

The results from the various techniques which we have applied so far
to simple models agree well with each other. This encourages us to
proceed a bit further and look at the
three-dimensional model corresponding to CsCuCl$_3$. This
means we consider the Hamiltonian (\ref{hamOp}) where the nearest
neighbour pairs $\langle i,j \rangle$ will now be those of the
two-dimensional triangular lattice.

\subsection{Ising expansions}

We start with Ising expansions.
The structure of the state with $\Mag = 1/3$ and the lowest
single-spin excitations above it is readily inferred from the
treatment of three coupled chains and the two-dimensional triangular
lattice \cite{twodim}. It is then straightforward to obtain fourth-order
series for these excitation energies\footnote{This can
probably be pushed a bit further. Higher-order series have been
computed numerically e.g.\ for the Heisenberg model on
three-dimensional cubic lattices \cite{WOH}.}. The explicit
series can be found in appendix A.
As before, these results correspond to the boundaries of the
$\Mag = 1/3$ plateau for sufficiently small $\Delta$.
At large values of $\Delta$, the associated transitions should be first
order such that one would have to look at other excitations to determine
these boundaries.

If we insert the value $\Jp = \abs{J}/3$
into the raw fourth-order series (\ref{hc3D1},\ref{hc3D2}) we find that
the plateau closes at $\Delta_c = 0.939$. Using instead
$\Jp = \abs{J}/6$ (which is appropriate for CsCuCl$_3$), the
ending point shifts to $\Delta_c = 0.948$. Both values are slightly
smaller than those obtained for the three-chain model with
the same order. However, in this region the actual values should
not be taken too seriously, {\it i.e.}\ higher orders should still
important and numerical data will indicate that the true $\Delta_c$
is probably larger. The main conclusion obtained from the Ising series
is that the $\Mag = 1/3$ plateau closes in the region $\Delta_c \approx 1$,
probably somewhere slightly below one in the region of XY-type anisotropies.

\subsection{Exact diagonalization}

\begin{figure}[t]
\centerline{\psfig{figure=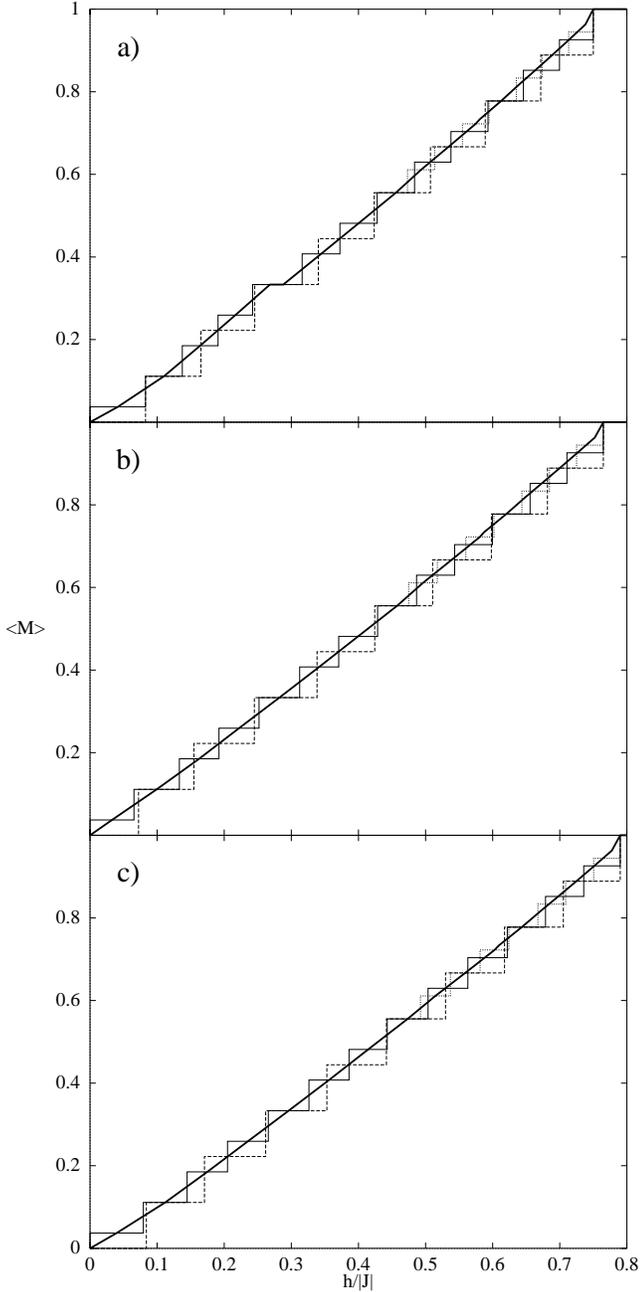,width=\columnwidth}}
\smallskip
\caption{Magnetization curves of $3 \times 3 \times L$ clusters with
a) $\Jp/\abs{J} = 1/6$, $\Delta = 1$;
b) $\Jp/\abs{J} = 1/6$, $\Delta = 0.97$;
c) $\Jp/\abs{J} = 0.172882193$, $\Delta_1 = 0.98789825$, $\Delta_2 = 1$.
The number of layers is $L=2$ (dashed line), $L=3$
(thin full line) and $L=4$ (dotted line).
The bold full line is an extrapolation to $L = \infty$.
\label{3dMag}
}
\end{figure}

To locate the ending point more accurately and to see if a jump
in the magnetization curve develops, we have also numerically computed
magnetization curves on three-dimensional $3 \times 3 \times L$
clusters. Some magnetization curves are shown in Fig.\ \ref{3dMag}.
Two criteria can be used to discuss the magnetization process in the
region $\Mag = 1/3$:
\begin{enumerate}
\item If the magnetization curve passes smoothly through $\Mag = 1/3$,
  the width of the finite-size plateau at $\Mag = 1/3$ should scale
  as $1/L$ (with $L$ the number of stacked segments of the triangular
  lattice).
\item For a smooth curve at magnetization $\Mag$, the neighbouring
  steps of the finite-size magnetization curves should have the same
  width. If a plateau is present, the step with magnetization $\Mag$
  should be broader than its neighbours. A jump at sufficiently large
  system sizes should in contrast be indicated by a step that is
  smaller than its neighbours at small system sizes.
\end{enumerate}
A remark is in order before we apply these criteria to our data.
Although we expect to obtain a reasonable approximation to the
thermodynamic limit despite the small linear size of the systems
accessible to us, one can clearly not rule out that the behaviour
would change on larger systems. Our conclusions should therefore
be taken as a probable possibility. 

For $\Jp/\abs{J} = 1/6$ and $\Delta = 1$, both criteria indicate
the presence of a small plateau with $\Mag = 1/3$. Instead for
$\Jp/\abs{J} = 1/6$ and $\Delta = 0.97$ both criteria applied to
$L=2$ and $L=3$ point to a smooth magnetization curve in the
region $\Mag = 1/3$. This suggests $1 > \Delta_c > 0.97$ which is
slightly larger than that found from the fourth-order Ising expansions.

Since we did not find evidence for a jump so far, we also performed
diagonalizations for the parameters $\Jp/\abs{J} = 0.172882193$,
$\Delta_1 = 0.98789825$, $\Delta_2 = 1$, since for these precise
parameters a jump was found in \cite{NiSh} using first-order spinwave theory.
The finite-size magnetization curves for these parameters are shown in
Fig.\ \ref{3dMag}c). Again, both criteria discussed above are compatible
with a smooth curve at $\Mag = 1/3$, {\it i.e.}\ neither a plateau nor
a jump seems to be present. However, on a system with volume
$27 = 3 \times 3 \times 3$, we would expect some indication for
a jump of size $\delta \Mag \approx 0.04$ as estimated in \cite{NiSh}
or even $\delta \Mag \approx 0.01$ as observed experimentally \cite{HTM}. 
One may attribute the absence of such signals for a jump to finite-size
effects beyond the discretization of the magnetization curve at a volume
with $27$ spins. However, it can be seen at larger $\Mag$ that increasing
the number of layers $L$ leads only to small finite-size effects.
On the other hand, also the approach of \cite{NiSh} is based on several
approximations and is therefore not guaranteed to describe the
situation with spin $1/2$ accurately.

To conclude this section, let us briefly comment on the utility of
order parameters which one might suspect to be useful for characterizing
the transition. First, one could check numerically that the correct
spin structures have been used in \cite{NiSh}. However, we do believe
that the spin structures are indeed the umbrella-type and coplanar
configurations in the low- and high-field regions, respectively.
In fact, this has been reliably established via a mapping to a gas
of hard-core bosons \cite{NSbose} for magnetic fields close to the
saturation value. The question which we raised above regards the
nature of the transition which occurs at intermediate fields,
{\it i.e.}\ if it is second order, first order or occurs e.g.\ via
an intermediate phase. However, we do not expect much additional
insight into the nature of the transition from a numerical determination
of order parameters, mainly because of finite-size effects such as the
discretization of the magnetization axis. For this purpose the 
magnetization $\Mag$ (which we have discussed above) might even
be somewhat better suited although it is not an order parameter.

\section{Conclusions}

We have theoretically observed a plateau with $\Mag = 1/3$
in a frustrated triangular magnet. For applications to CsCuCl$_3$,
it is important to observe that the width of this plateau depends
on several factors.

Consider for example a single triangle. If there are only three
spin-$1/2$ spins, the magnetization can obviously only have
the values $\Mag = \pm 1/3$ and $\pm 1$. The plateau with
$\Mag = 1/3$ survives (at least for $\Delta = 1$) the ferromagnetic
stacking of infinitely many triangles, but the interaction reduces its width.
A similar phenomenon occurs if one instead takes the
limit of a two-dimensional triangular lattice in the plane.
Also here, the plateau survives the thermodynamic limit
for $\Delta \approx 1$ \cite{MiNi,twodim}, but again gets narrower.

A further factor is the XXZ-anisotropy $\Delta$. For a magnetic
field along the $c$-axis, the plateau with $\Mag = 1/3$
survives only a small amount of XY-like anisotropy before it
disappears. The same observation has already been made for
the triangular lattice antiferromagnet \cite{MiNi,twodim}.
If the magnetic field is applied in the triangular
plane perpendicular to the $c$-axis, an opposite behaviour is
known from the triangular lattice antiferromagnet \cite{SuMa}:
Here one observes a clear plateau-like feature in the region
$\Mag = 1/3$ even in the XY case $\Delta = 0$.

If one combines these observations with the parameters which are
appropriate for the description of CsCuCl$_3$, one understands the
absence of a plateau for a magnetic field in the direction of the
$c$-axis and that a plateau-like feature  only occurs if the magnetic
field is applied in the plane \cite{HTM}.

In contrast, we failed to reproduce the jump observed experimentally
\cite{HTM} for a magnetic field along the $c$-axis with the model
(\ref{hamOp}) in the appropriate parameter region.
While we observe several first-order transitions for $\Delta > 1$,
we have no evidence for first-order transitions for $\Delta < 1$.
Actually, our results for the three-chain model as well as similar
ones for the two-dimensional triangular lattice \cite{twodim} provide
quite strong evidence against such a jump for spin $1/2$ and $\Delta < 1$.
However, a first-order spinwave analysis \cite{NiSh} predicted 
a jump in the magnetization curve of the Hamiltonian (\ref{hamOp})
with $\Delta_1 < 1$. Since it may be possible that this jump cannot
occur in less than three dimensions, we have also investigated the
three-dimensional spin-$1/2$ model numerically. Using precisely the
same parameters as in \cite{NiSh}, we did not find any evidence for
a jump on $3 \times 3 \times L$ clusters either. This discrepancy
is puzzling, in particular since the weak-coupling analysis shows
that weakly coupled ferromagnetic chains are close to a classical
situation (though in two rather than in three dimensions\footnote{
More precisely, each ferromagnetic chain can be regarded as one
single spin with large $S$. It might therefore actually be interesting
to try to treat our second-order expression in $\Jp$ in three dimensions
using spinwave techniques.}).
While we cannot fully exclude that one would find a jump in the spin-$1/2$
model at larger system sizes, also the discussion in \cite{NiSh} is based
on several approximations. We therefore believe that the issue of the
experimentally observed jump deserves further theoretical attention.
One possibility is also that this jump is caused by effects
which are not incorporated into the model (\ref{hamOp}), such as
elastic response to the external magnetic field. The anomalies observed
in a recent soundwave propagation experiment \cite{WZSL}
may be a further indication for the necessity of such extensions.

In summary, we have shown that a three-chain model describes the
low-temperature magnetization process of CsCuCl$_3$ almost as well
as the full three-dimensional model. The three-chain model has the
advantage of being simpler to analyze.
We have concentrated on zero (or small) temperature, but there is also
a large amount of experimental data on the $h$-$T$ phase diagram (see
\cite{WWWLS,SKS,SWSZKL} for a small selection of recent results)
which are waiting to be discussed in the framework of microscopic
models. The three-chain model which we have discussed in the present
paper may be a useful starting point since one should be able to treat
it with a finite-temperature variant \cite{WaXia,Shibata,DMRGbook} of
the DMRG procedure.

{}From a technical point of view, we have shown that the Hamiltonian
(\ref{hamOp}) with spin $1/2$ can be analyzed by several methods.
Each of them has the potential of being pushed further to obtain
more accurate results. This applies in particular to the weak-coupling
approach: The second order diverges for the three-chain model as
$L \to \infty$, but it remains to be investigated if it
converges in three dimensions.
The computation of other quantities than just the magnetization is
also straightforward with each of the methods used in the present paper.

\begin{acknowledgement}

Useful discussions with D.C.\ Cabra, T.\ Nikuni, I.\ Peschel, P.\ Pujol,
U.\ Schotte, H.\ Tanaka and M.E.\ Zhitomirsky are gratefully acknowledged.
We are indebted to the Max-Planck-Institut f\"ur Mathematik, Bonn-Beuel
and the C4 cluster of the ETH for allocation of CPU time.

\end{acknowledgement}

\appendix

\section{Ising series}

In this appendix we present explicit series in the Ising anisotropy
$\Delta^{-1}$.

For the $\Mag = 1/3$ plateau in the {\it three-chain model}, one
finds the following sixth-order series for the gap of a single spin
flipped up
\bea
h_{c_2} &=&  \Delta \left (\Jp-J\right ) +J
+{\Jp^{2}} \Biggl\{ {1 \over 4 J} \Delta^{-1}
+{5 \Jp-2 J \over 16 J^{2}} \Delta^{-2} \nn \\
&+& {164 J^{2}\Jp^{2}-258 J^{3}\Jp+276 J^{4}
  -93 \Jp^{3}J+19 \Jp^{4} \over
  192 J^{3}\left (\Jp-J\right )\left (\Jp-2 J\right )}
  \Delta^{-3} \nn \\
&+& {\Delta^{-4} \over 2304 J^{4}\left (\Jp-J\right )^{2}
  \left (\Jp-2 J \right )^{2} }
  \left( 19050 \Jp^{3}J^{4} \right. \nn \\
&& \quad -8234 \Jp^{4}J^{3} +3297 \Jp^{5}J^{2}
   -27548 \Jp^{2}J^{5} \nn \\
&& \quad \left. +20232 \Jp J^{6}
-1136 \Jp^{6}J+171 \Jp^{7}-6480 J^{7} \right) \nn \\
&+& {\Delta^{-5} \over 55296 \left (\Jp- 2 J\right )^{3}J^{5}
\left (\Jp-3 J\right )\left (\Jp-4 J \right )
\left (\Jp-J\right )^{3} } \nn \\
&& \ \left(62217652 \Jp^{4}J^{8}- 35026088 \Jp^{5}J^{7} \right. \nn \\
&& \quad -76513096 \Jp^{3}J^{9} +65015952 {\Jp}^{2}J^{10} \nn \\
&& \quad +14995748 \Jp^{6}J^{6} -7562886 \Jp^{7}J^{5} \nn \\
&& \quad +8581 \Jp^{12} +4998735 \Jp^{8}J^{4} -2554610 \Jp^{9}J^{3} \nn \\
&& \quad +783380 \Jp^{10}J^{2}-128008 \Jp^{11}J \nn \\
&& \quad \left. -35276256 J^{11}\Jp + 9134208 J^{12} \right)
\Biggr\} \nn \\
&&+ \Order(\Delta^{-6}) \, . \label{hc2ThreeChain}
\eea
For an excitation with a single spin flipped down one finds instead
the following sixth-order series
\bea
h_{c_1} &=& \Delta J - J + {\Jp \over 2}
- \Jp^2 \Biggl\{ {2 \Jp -3 J \over 16 J^{2}} \Delta^{-2} \nn \\
&+&{47 J^{2}\Jp^{2}-39 J^{3}\Jp+30 J^{4}-25 \Jp^{3}J+5
  \Jp^{4} \over 64 J^{3}\left (\Jp-J\right )
  \left (\Jp-2 J\right )} \Delta^{-3} \nn \\
&+&{\Delta^{-4} \over
  4608 J^{4}\left (\Jp- J\right )^{2}\left (\Jp-2 J\right )^{2}}
  \left(24218 \Jp^{3}J^{4} \right. \nn \\
&& \quad -17776 \Jp^{4}J^{3}
   +10901 \Jp^{5}J^{2}-30028 \Jp^{2}J^{5} \nn \\
&& \quad \left. + 24264 \Jp J^{6}
    -4054 \Jp^{6}J+611 \Jp^{7} -8784 J^{7} \right) \nn \\
&+&{\Delta^{-5} \over
 110592 \left (\Jp -J\right )^{3}\left (\Jp-4 J\right )
  \left (\Jp-3 J \right )J^{5}} \nn \\
&& \ \times {1 \over \left (\Jp-2 J\right )^{3}
  \left (2 \Jp -3 J\right )} \left(
    468904952 \Jp^{5}J^{8} \right. \nn \\
&& \quad -400904412 \Jp^{6}J^{7} -488731468 \Jp^{4}J^{9} \nn \\
&& \quad +415953144 \Jp^{3}J^{10} -263997936 \Jp^{2}J^{11} \nn \\
&& \quad +309762362 \Jp^{7}J^{6} -199502115 \Jp^{8}J^{5} \nn \\
&& \quad -23421312 J^{13} +51214 \Jp^{13}+111406752 J^{12}\Jp \nn \\
&& \quad +97003020 \Jp^{9}J^{4}-32952968 \Jp^{10}J^{3} \nn \\
&& \quad \left.  +7259020 \Jp^{11}J^{2}-923565 \Jp^{12}J
  \right) \Biggr\} \nn \\
&& + \Order(\Delta^{-6}) \, .
\label{hc1ThreeChain}
\eea

In a similar way, one obtains the following fourth-order series
for the single-spin excitation above $\Mag = 1/3$ in the
{\it three-dimensional model}
\bea
h_{c_1} &=& J \Delta + {3 \Jp \over 2}-J \nn \\
&+& {3 \Jp^{3}
\left (9 J{\Jp}-10 J^{2}+10 {\Jp}^{2}\right ) \over 8 \Delta
\left ({\Jp}-J\right )\left (3 {\Jp}-2 J\right )\left (2 {\Jp}-J
\right )\left ({\Jp}-2 J\right )} \nn \\
&-& {3 \Jp^2 \over 32 \Delta^{2}
   \left (\Jp-2 J\right )^{2} \left (2 \Jp-J\right )^{2}
   \left (3 \Jp-2 J\right ) \left (\Jp-J\right )^{2} } \nn \\
&& \ \left (292 \Jp^6-900 J \Jp^5 +1203 J^{2}{\Jp}^{4}
    -1222 J^{3} \Jp^{3} \right. \nn \\
&& \qquad \left.
  +896 J^{4}{\Jp}^{2} -344 J^{5} \Jp +48 J^{6}\right ) \nn \\
&+& {\Jp^2 \over 256 \Delta^3}
\left (3696796800 {\Jp}^{18} -5988208896 {\Jp}^{3}J^{15} \right. \nn \\
&& \ +4368432498318 {\Jp}^{12}J^{6} -3140479722337{\Jp}^{13}J^{5} \nn \\
&& \ -3154044909140 {\Jp}^{9}J^{9} -41660867520 {\Jp}^{17}J \nn \\
&& \ +30196054016 {\Jp}^{4}J^{14} +221631078840 {\Jp}^{16}J^{2} \nn \\
&& \ -117183492672 {\Jp}^{5}J^{13} -742570973868 {\Jp}^{15}J^{3} \nn \\
&& \ +362869230400 {\Jp}^{6}J^{12} +1760559743358 {\Jp}^{14}J^{4} \nn \\
&& \ -912414730048 {\Jp}^{7}J^{11} +1875316243840 {\Jp}^{8}J^{10} \nn \\
&& \ +4333067026982 {\Jp}^{10}J^{8} -4842206323349 {\Jp}^{11}J^{7} \nn \\
&& \ \left. +854889984 {\Jp}^{2}J^{16}
          -77524992 J^{17}{\Jp} +3317760 J^{18}\right ) \nn \\
&& \ \times \biggl\{
\left ({\Jp}-J\right )^{3}\left (3 { \Jp}-2 J\right )^{3}
\left (2 {\Jp}-J\right )^{3} \left ({\Jp}-2 J\right )^{3} \nn \\
&& \qquad \left (3 {\Jp}-4 J\right ) \left (5 {\Jp}-4 J\right )
\left (4 {\Jp}-3 J\right ) \left (5 {\Jp}-3 J\right ) \nn \\
&& \qquad \left (4 {\Jp}-J\right )\left (3 {\Jp}-J\right )
\left(5 {\Jp}-2 J\right ) \biggr\}^{-1} \nn \\
&+& \Order(\Delta^{-4}) \, ,
\label{hc3D1}
\eea
and
\bea
h_{c_2} &=& \left (3 {\Jp}-J\right )\Delta+J-{3 \Jp^{2}\left (6
J^{2}-11 J{\Jp}+10 {\Jp}^{2}\right ) \over 4 \Delta \left ({\Jp}-J\right )
\left (2 {\Jp}-J\right )\left ({\Jp}-2 J\right )} \nn \\
&&+ {3 \Jp^{2}\left (50 \Jp^{3}-37 J\Jp^{2}-12 J^{2}\Jp
 +12 J^{3}\right ) \over 16 \Delta^2 \left (\Jp- 2 J\right )
\left (2 \Jp-J\right) \left(\Jp-J\right )^{2}} \nn \\
&&+ {\Jp^{2} \over 64 \Delta^{3}}
\left(13530240 {\Jp}^{15}+80661600 {\Jp}^{2}J^{13} \right. \nn \\
&& \qquad -5086562914 {\Jp}^{9}J^{6} +1224415456 {\Jp}^{4}J^{11} \nn \\
&& \qquad -2716474016 {\Jp}^{5}J^{10}+4432214446 {\Jp}^{6}J^{9} \nn \\
&& \qquad -387107136 {\Jp}^{3}J^{12}-5597333776 {\Jp}^{7}J^{8} \nn \\
&& \qquad +5797552217 {\Jp}^{8}J^{7} +3611625955 {\Jp}^{10}J^{5} \nn \\
&& \qquad -1800747488 \Jp^{11} J^{4} +447946876 {\Jp}^{12}J^{3} \nn \\
&& \qquad + 60036504 {\Jp}^{13}J^{2} -70410816 {\Jp}^{14}J \nn \\
&& \qquad \left. -9897984 J^{14}{\Jp}+539136 J^{15} \right)\nn \\
&& \ \times \biggl\{
\left ({\Jp}-2 J\right )^{3}\left ({\Jp
}-J\right )^{3}\left (3 {\Jp}-2 J\right )\left (3 {\Jp}-4 J
\right ) \nn \\
&&\qquad \left (5 {\Jp}-4 J\right )\left (4 {\Jp}-3 J\right )
\left (2 {\Jp}-J\right )^{3}\left (3 {\Jp}-J\right ) \nn \\
&&\qquad \left (5{\Jp}-3 J\right )\left (4 {\Jp}-J\right ) \biggr\}^{-1} \nn \\
&&+ \Order(\Delta^{-4}) \, .
\label{hc3D2}
\eea
The results (\ref{hc3D1},\ref{hc3D2}) are consistent with earlier ones.
Firstly, for $J = 0$, one recovers the series for the triangular lattice
\cite{twodim}. For general $J < 0$ also (\ref{hc2ThreeChain}) and
(\ref{hc1ThreeChain}) match with (\ref{hc3D2}) and (\ref{hc3D1}),
respectively up to first order in $\Jp$ (one just has to rescale $\Jp$ by
a factor $3$ to account for the different coordination
number).

\section{Details of the DMRG procedure}

Following the standard infinite-size algorithm of the DMRG
method~\cite{White}, we enlarge the chain in each iteration step by
two triangles (in total six spin-$1/2$ sites), {\em i.e.} by
$8\times 8=64$ states. This is different from~\cite{tandon} where the
totally antiferromagnetic case of the Hamiltonian (\ref{hamOp}) has
been treated. There, the chain was built up by adding a total of only
two spin-$1/2$ sites in each step, so that it took
three DMRG steps to add two complete triangles to the original
ladder. In this procedure more truncation operations of the Hilbert
space had to be performed than in our calculation. But by adding
eight-state sites in each step the number of basis states discarded in
the truncation process is larger than in the case where only two-state
sites are added. As far as we know it has not yet been tested which
procedure is more favourable.

In order to limit the computational effort, we applied only the
infinite-system algorithm even though more accurate results could probably
be obtained when one would use in addition the finite-system algorithm.
To calculate the boundaries of the magnetization plateau we have used spin
conservation and performed three different calculations to target each of the
groundstates with magnetization $\Mag=1/3-1/S_{\rm tot}$, $1/3$, $1/3+1/S_{\rm tot}$
separately. The difference between the groundstate energy in the sector
$\Mag=1/3$ and the energies in the neighbouring sectors
then gives the magnetic field:
\begin{eqnarray}
h_{c1} & = & \textstyle
 E_0\left(\Mag=\frac{1}{3}+\frac{1}{S_{\rm tot}},L\right)-
 E_0\left(\Mag=\frac{1}{3},L\right) \qquad \\ 
h_{c2} & = & \textstyle E_0\left(\Mag=\frac{1}{3},L\right)
           - E_0\left(\Mag=\frac{1}{3}-\frac{1}{S_{\rm tot}},L\right).
\end{eqnarray}

\begin{figure}[ht]
\centerline{\psfig{figure=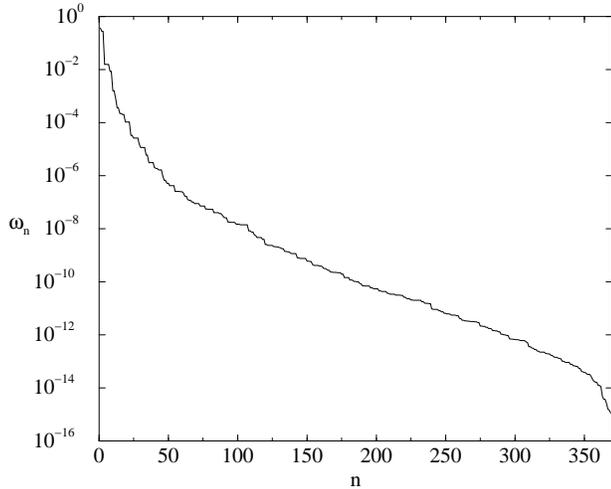,width=0.9\columnwidth}}
\smallskip
\caption{\label{fig_spec_rho} Eigenvalue spectrum of the reduced
density matrix of a ladder with 24 spins ($L=8$), calculated with $64$
kept states in the subspace $\Mag=1/3$. The couplings are $J=-1$,
$J'=1/3$ and $\Delta=1$}
\end{figure}

Fig.\ \ref{fig_spec_rho} shows a typical spectrum of the eigenvalues
of the density matrix for such a calculation.

The eigenvalues drop exponentially as is known from other quantum spin
chains\footnote{Although the number of applications of the DMRG has
increased in the last few years, only a small amount of eigenvalue spectra
of the used density matrices can be found in the literature.
Some spectra are discussed in \cite{White,LePa,KauPe,PeKauLe,OHA}.}
and thus gives rise to a very small truncation
error. On the other hand, we found that at least 160 states of the
density matrix, {\it i.e.}\ 1280 states in one part of the chain had to
be kept to obtain reliable results up to three decimal places in the
energy. The eigenvalues of the density matrix lie about three to four
orders of magnitude above the eigenvalues in the corresponding
spectrum \cite{Tpriv} of the totally antiferromagnetic case studied
in \cite{tandon}. This is one reason for the lower accuracy of our DMRG
calculations. There is also a big difference between the truncation error
and the real error in the groundstate energy which indicates that the
structure of the groundstate could be approximated better by applying the
finite-system algorithm.

\end{document}